\documentclass[usenatbib]{basi}
\usepackage[british]{babel}
\usepackage[varg]{txfonts}
\usepackage{rotating}
\usepackage{dcolumn}
\begin{document}
\title{``EIT waves" and coronal mass ejections}
\author[P.~F.~Chen \& C. Fang]%
       {P.~F.~Chen$^{1,2}$\thanks{email: \texttt{chenpf@nju.edu.cn}} and
       C.~Fang$^{1,2}$\\
       $^1$Department of Astronomy, Nanjing University, Nanjing 210093,
        China\\
       $^2$Key Lab of Modern Astron. \& Astrophys. (Ministry of
        Education), Nanjing University, China}

\pubyear{2011}
\volume{00}
\pagerange{\pageref{firstpage}--\pageref{lastpage}}

\date{Received \today}

\maketitle
\label{firstpage}

\begin{abstract}
Coronal ``EIT waves" appear as EUV bright fronts propagating across a
significant part of the solar disk. The intriguing phenomenon provoked
continuing debates on their nature and their relation with coronal mass
ejections (CMEs). In this paper, we first summarize all the observational
features of ``EIT waves", which should be accounted for by any successful
model. The theoretical models constructed during the past 10 years are then
reviewed. Finally, the implication of the ``EIT wave" research to the
understanding of CMEs is discussed. The necessity is pointed out to revisit
the nature of CME frontal loop.
\end{abstract}

\begin{keywords}
   waves -- Sun: coronal mass ejections (CMEs) -- Sun: corona
\end{keywords}

\section{Introduction}\label{s:intro}

Coronal mass ejections (CMEs) are observed as enhanced brightness propagating
out from the low solar corona. A typical CME consists of a bright frontal loop,
a bright core, and a cavity in between. Since the discovery in early 1970s, CMEs
have been studied extensively. As the largest-scale eruptive phenomenon in the
solar atmosphere, they were verified to be the major driver of the disastrous
space weather environment. Therefore, CMEs have received continuous attention in
the whole community, and various efforts were devoted to the investigations on
them and their relations with all other accompanied phenomena, such as solar
flares, filament eruptions, radio bursts, particle accelerations, and so on.
However, a fundamental question still remains, i.e., what is the nature of CMEs?

When a pattern is observed to move, there are three possibilities. First, it can
be a wave, such as the surface wave on a lake. Second, it can be a mass motion,
such as the erupting prominence. The third possibility, which was often
neglected, is the apparent motion, such as the flare ribbon separation, which
is neither a wave nor a mass motion. It is vital to combine imaging and
spectroscopic observations to distinguish among these three possibilities, which
is however often hard to do. In terms of CMEs, they were considered to be
fast-mode magnetohydrodynamic (MHD) waves driven by solar flares in the 1970s.
Such an idea was discarded soon since it contradicts with many observational
features. Since then, CMEs have been taken for granted to be mass motions, and
the measured velocity based on the white-light coronagraph observations has been
considered to be the bulk velocity projected to the plane of the sky.

The bright core of a CME can be identified to be the erupting filament (or
prominence), whose propagation is definitely a mass motion. However, the
propagation of the CME frontal loop is not so obvious \citep{chen09a}. It might
be thought that spectroscopic measurements can easily clarify such an issue.
However, CMEs and their dynamics are better resolved for the events that
propagate not far from the plane of the sky, whereas spectroscopic
measurements are valid for the CMEs whose propagation significantly
deviates from the plane of the sky. The very rare imaging and
spectroscopic observations of several halo CMEs indicated that the
propagation of the CME frontal loop is not bulk motion, and the plasma
velocity is several times smaller than the apparent velocity measured
in the white-light images \citep{ciar06}, i.e., similar to a wave, there
is mass motion, but the mass motion is several times slower than the
propagation of the bright fronts.

As seen above, the nature of CME frontal loop is not so well established
as most people have presumed. Its nature deserves deeper
investigations. Just as our understanding on CMEs benefited a lot from the
studies on the CME-related phenomena like flares and radio bursts, the
nature of the CME frontal loop might also be hidden in the observational
and modeling studies of CME-related phenomena, in particular, EIT waves.
In this paper, we give a brief review on EIT waves, and explicate how the
EIT wave modelings can shed light on our understanding of CMEs.

\section{Observations of EIT waves}\label{s:obs}

When talking about EIT waves, we have to mention another wave phenomenon,
i.e., Moreton waves. More than 50 years ago, \citet{more60} discovered a
dark front in the H$\alpha$ red wing (or a bright front in the H$\alpha$
blue wing) images, propagating out for a distance on the order of $5\times
10^5$ km from some big flares, with a velocity ranging from 500 to 2000 km
s$^{-1}$. It was later called Moreton waves. H$\alpha$ line is formed in
the chromosphere, therefore, Moreton wave is a chromospheric phenomenon.
However, considering that the Alfv\'en velocity in the quiet chromosphere
is typically 100 km s$^{-1}$, Moreton wave cannot be a wave of chromospheric
origin, since its fast speed would otherwise imply a strong shock wave (with
a Mach number of 5-20), which cannot sustain for a long distance. Such a
puzzle was solved later by \citet{uchi68}, who proposed that Moreton waves
are due to a fast-mode MHD shock wave in the corona, sweeping the
chromosphere to produce the apparent propagation of Moreton wave fronts.
Since the fast-mode wave speed in the corona is several times higher than
in the chromosphere, the shock wave is not necessarily very strong, so it
can propagate for a long distance. Such a model predicts that there should
be a fast-mode wave in the corona coming out from a flare site with a
velocity of 500--2000 km s$^{-1}$, which should be detected in X-ray and
EUV wavelengths. The detection of the coronal fast-mode wave was extremely
rare, with a probable candidate found by \citet{neup89} and several other
events studied by \citet{khan02}, \citet{huds03}, and \citet{naru04}. The
wave speeds in these events are in the typical velocity range of Moreton waves.

\begin{figure}
\centerline{\includegraphics[width=13cm]{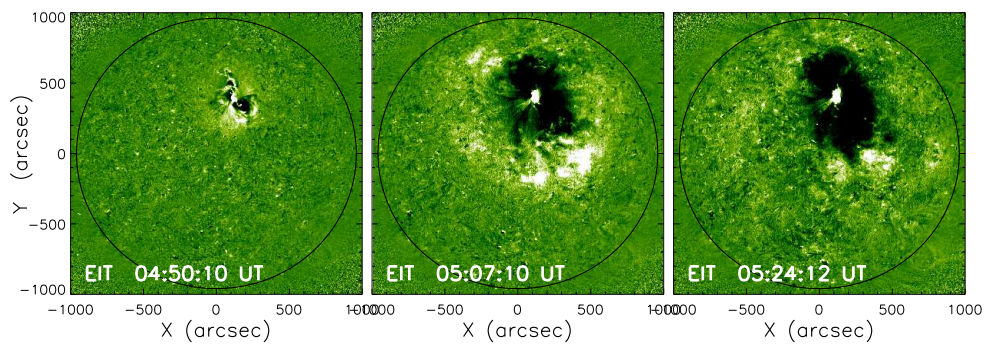}}
\caption{EIT 195~{\AA}\ base difference images showing the evolution of the
	most famous EIT wave event on 1997 May 12 \citep[from][]{chen11}.
        \label{fig:eit}}
\end{figure}

After the launch of the {\it Solar and Heliospheric Observatory} ({\it SOHO})
spacecraft, one of its payload, EUV Imaging Telescopes \citep[EIT;][]{dela95},
began to monitor the full solar disk in 4 EUV channels, with a cadence of
$\sim$15 min for the 195{\AA} channel. Using the running difference technique,
\citet{thom98} found that a large-scale wave, with bright fronts immediately
followed by extending dimmings, propagates out from the flaring site, with a
velocity of 250 km s$^{-1}$, as illustrated by Fig. \ref{fig:eit}. They were
named ``EIT waves" after the telescope. Such an
interesting phenomenon sparked wide interest, as well as controversies,
in the community. It is hotly debated whether EIT waves are the
long-awaited coronal counterparts of H$\alpha$ Moreton waves or not. In
this section, we summarize the typical observational features of ``EIT
waves". It is expected that any successful model should explain all
these characteristics.

{\bf (1) The velocity}

\citet{klas00} carried out a statistical study on the EIT wave velocity
based on the EIT observations in 1997, and it was found that the velocity
varies from 138 to 465 km s$^{-1}$, with an average of 271 km s$^{-1}$.
With a higher cadence of 2.5 min, the Extreme Ultraviolet Imager (EUVI)
on board the {\it STEREO} spacecraft revealed that the EIT wave velocity
can be as small as $\sim 10$ km s$^{-1}$ \citep{zhuk09}, which is even
much smaller than the sound speed in the corona. \citet{long08} pointed
out that the low cadence observations by {\it SOHO}/EIT would underestimate
the EIT wave velocity. However, we note that a fair argument is that
low-cadence observations would underestimate the peak velocity and
overestimate the trough velocity when the EIT wave speed changes with time.

Furthermore, \citet{klas00} found that the EIT wave velocity is
generally $>3$ times slower than the associated type II radio bursts,
and the velocities of these two phenomena have no any correlation. Note
that type II radio bursts have been well established to be due to the
fast-mode shock wave in the corona.

It is also noticed that several authors have shown that EIT waves
accelerate when they move from the proximity of source active region
to the quiet region, and then decelerate \citep{long08, zhuk09, yang10,
liuni10}.

{\bf (2) Stationary fronts}

EIT waves were found to finally stop somewhere, e.g., 
\citet{thom99} found that EIT waves stop at the boundary of
coronal holes, and \citet{dela99} revealed that a propagating EIT wave
stopped at the footpoints of coronal magnetic separatrix. These two
features are consistent since the boundary of coronal holes is also a
magnetic separatrix.

\citet{gopa09} analyzed {\it STEREO}/EUVI running difference images and
claimed that an EIT wave was bounced back as it hit the boundary of a
low-altitude coronal hole. On the contrary, \citet{attr10} studied that
same event with the base difference images and argued that the 
reflecting EIT waves in \citet{gopa09} might be an illusion, and the
EIT wave actually stopped near the coronal hole boundary.

{\bf (3) Relation with solar flares}

\citet{cliv05} pointed out that half of the EIT waves are associated
with weak flares, such as {\it GOES} A- or B-class events, posing doubt
on whether the pressure pulse in flares can generate the global-scale
EIT waves. Following this line of thought, \citet{chen06} did a test to
examine whether solar flares alone can generate EIT waves. The results
indicate that, without CMEs, even M- and X-class flares cannot
produce EIT waves. Note that occasionally people claim that a flare
without a CME was associated with an EIT wave. It is presumably that the CME
was missed by the coronagraph due to low Thomson-scattering \citep{zhan10}.

{\bf (4) Relation with CMEs}

Based on the statistical investigations, \citet{bies02} concluded that
EIT waves are intimately related to CMEs, rather than flares. The
test by \citet{chen06} also indicates that no matter the associated 
flare is strong or weak, EIT waves can be observed only if a CME is
present.

Nowadays, it is widely accepted that EIT waves are directly linked to CMEs.
However, there still exists a dispute on the spatial relation between EIT waves
and CMEs. \citet{chen09a} and \citet{dai10} found that the EIT wave front is
cospatial with the CME frontal loop, whereas \citet{pats09} and \citet{vero10}
argued that the EIT wave front is further away from the CME frontal loop. This
issue should be clarified.

{\bf (5) Other features}

(a) \citet{attrha07} found that as the EIT wave front propagates outward, the
location of the peak intensity rotates apparently in the same direction 
(clockwise or anti-clockwise) as the erupting filament;

(b) \citet{harr03} found that the Doppler velocity is negligible in
	the EIT wave fronts and significant in the extending dimmings that
	are immediately behind the EIT wave fronts;

(c) \citet{yang10} examined the relation between the EIT wave velocity
	and the local magnetic field strength. They found that the two
	quantities often show a negative correlation, which does not favor the
        fast-mode wave model for EIT waves;

(d) There exists significant line broadening behind the EIT wave front
	\citep{chen10}.

\section{Modelings of EIT waves}

In order to interpret the intriguing phenomenon, several models have been
proposed so far \citep[see][for reviews]{willat09, warm10, galllo11, chen11}.
Here, we briefly introduce several models. It is noted that EIT waves can
be applied to diagnose the coronal magnetic field. However, the results
critically depend on our understanding of EIT waves \citep{warm04, ball07,
chen09b}.

\subsection{Fast-mode wave model}\label{fast}

EIT waves were widely thought to be the coronal counterparts of H$\alpha$
Moreton waves, i.e., they are fast-mode waves in the corona \citep{wang00,
wu01, warm01, vrsn02, warm04, ball05, grec08, pomo08, vero08, gopa09, pats09,
muhr10}. In order to reconcile the large difference between Moreton waves
and EIT waves, \citet{wu01} and \citet{warm01} proposed that the fast-mode
wave speed decreases, say by $\sim$3 times, from the active region to the
quiet region. Similarly, \citet{grec11} suggested that the EIT wave velocity
profile fits the decelerating self-similar solutions very well. It is noted
that the finding of a remote filament winking implies that the Moreton wave
does not decelerate \citep{eto02}, the observations by {\it STEREO}/EUVI
also do not show decelerations \citep{ma09}.

It is noted that the popular fast-mode wave model can hardly explain many
features of EIT waves, such as their extremely low speed that is even smaller
than the sound speed, their stationary fronts, their cospatiality with CME
frontal loop, and so on.

\subsection{Successive field-line stretching model}

\begin{figure}
\centerline{\includegraphics[width=10.cm]{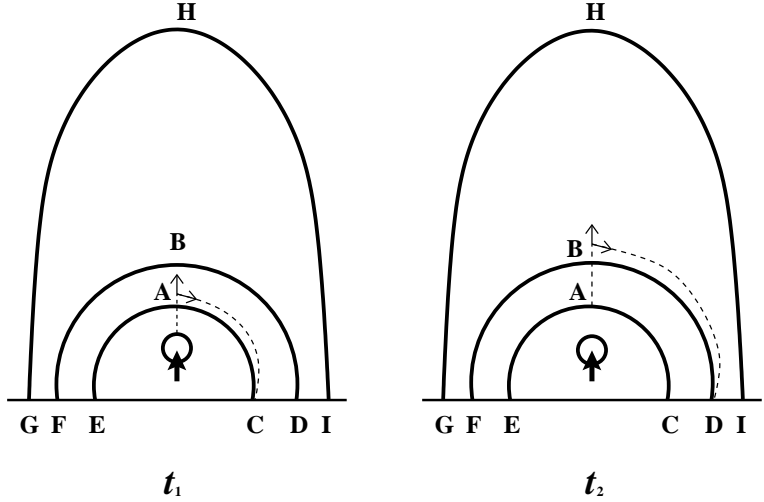}}
\caption{Schematics of the successive field-line stretching model for EIT
 waves \citep[from][]{chen05}. \label{fig:chen}}
\end{figure}

Inspired by the doubting of \citet{dela99} and  \citet{dela00}, \citet{chen02,
chen05} proposed that EIT waves are apparent motions of brightenings that are
generated by the compression as the magnetic field lines overlying the erupting
flux rope are pushed to stretch up successively. This model was deduced naturally
by realizing two facts: (1) All the field lines overlying the flux rope would be
stretched outward successively during CMEs; (2) For each field line, the
stretching starts from the top, and is then transferred down to the footpoints.
The formation of EIT waves in this model is illustrated in Fig. \ref{fig:chen},
which can be understood as follows: As the flux rope (the circle in the figure)
erupts, it pushes the first field line at point A, and then the perturbation
propagates to point C with the local fast-mode wave speed. At the same time, the
stretching propagates from point A to point B and then to point D with the local
fast-mode wave speed. Wherever the stretching comes, the local plasma is
compressed to form brightenings, i.e., EIT wave fronts. Therefore, the apparent
speed for the EIT wave to propagate from point C to point D is $v_{\mathrm{EIT}}
=CD/\Delta t$, with $\Delta t=\int_A^B1/v_f {\rm d}s+\int_B^D1/v_A {\rm d}s-
\int_A^C1/v_A {\rm d}s$, where $v_A$ is the Alfv\'en speed, and $v_f$ is the
fast-mode wave speed perpendicular to the field line, and the last two integrals
are along the field line shown in Fig. \ref{fig:chen}. If the field lines are 
semicircular, it is derived that the EIT wave speed is about 1/3 of the local
fast-mode wave speed. The erupting flux rope would also excite a piston-driven
shock wave, which straddles over the flux rope and extends down to the solar 
surface. Different from the EIT waves, the fast-mode shock wave propagates 
outward with a speed slightly larger than the local fast-mode wave speed.

This model predicts that the CME-driven (not flare-driven) shock wave is the
counterpart of H$\alpha$ Moreton wave, which runs ahead of the associated EIT
waves, with a speed of $\sim$3 times faster. \citet{harr03} found evidence of
a faster wave ahead of the EIT wave with the {\it TRACE} observations, and
recently, \citet{chenwu11} confirmed the the coexistence of a faster wave and an
EIT wave with the Solar Dynamics Observatory \citep[{\it SDO},][]{titlho06}
observations. In 3-dimensional MHD simulations, \citet{down11} also found that
a fast-mode wave runs ahead of the EIT wave.

\subsection{Successive reconnection model}

Noticing that the EIT wave fronts rotate apparently in the same direction as
the erupting filament, \citet{attrha07} also claimed that EIT waves should be
related to the magnetic rearrangement, rather than an MHD wave. 
They proposed a successive reconnection model,
i.e., EIT wave fronts are the footprint of the CME frontal loop, which
is formed due to successive magnetic reconnection between the expanding
core field lines and the small-scale opposite polarity loops. As more and more
field lines are pushed to stretch up, some of them may have a chance to
reconnect with neighboring loops \citep{cohe09}, it is a little hard to imagine
that this accounts for most of EIT wave fronts.

\subsection{Slow-mode (soliton) wave model}

Noticing that EIT waves generally keep single-pulse fronts and that the
EIT wave velocity is sometime smaller than the sound speed in the
corona, \citet{willde07} speculated that the EIT waves might be best
explained as a soliton-like phenomenon, say, a slow-mode solitary wave.
They stated that a solitary wave model can also explain other properties
of the EIT waves, such as their stable morphology, the non-linearity of
their density perturbations, the lack of a single representative
velocity, and their independence of Moreton waves. Such an idea requires
further quantitative modelings, which are not so straightforward in
2- or 3-dimensions \citep{willat09}.

\citet{wang09} performed 2-dimensional MHD numerical simulations of a flux rope
eruption, where they found that behind the piston-driven shock appear
velocity vortices and slow-mode shock waves. They interpret the vortices
and the slow-mode shock wave as the EIT waves, which are 40\% as fast
as the Moreton waves.

\subsection{Current shell model}

Through 3-dimensional MHD simulations, \citet{dela08} found that as a flux tube
erupts, an electric current shell is formed by the return currents of the
system, which separate the twisted flux tube from the surrounding fields.
Slightly different from their early idea of
magnetic rearrangement \citep{dela99}, they claim that this current shell
corresponds to the ``EIT waves''. They also revealed that the
current shell rotates, similar to the apparent rotation of the EIT wave
fronts found by \citet{podl05}. They emphasized the role of Joule
heating in the current shell in explaining the EIT wave brightenings,
which was not agreed by \citet{willat09}.

\section{Implications to the nature of CMEs}

\begin{figure}
\centerline{\includegraphics[width=13cm]{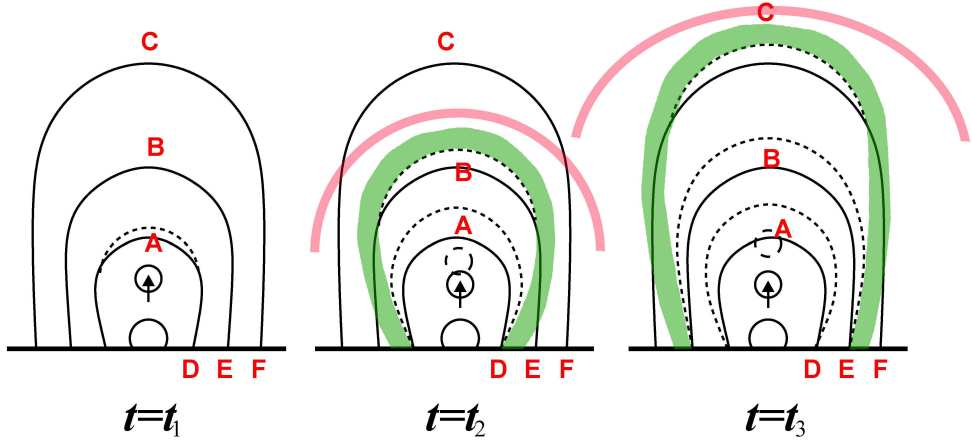}}
\caption{A schematic sketch of the formation mechanism of CME leading
loops, where the CME leading loop ({\it green}) are apparently-moving
density enhanced structure that is generated by the successive
stretching of magnetic field lines as the erupting core structure, e.g.,
a flux rope, continues to push the overlying field lines to expand
outward successively. The piston-driven shock is shown as pink lines
\citep[from][]{chen09a}. \label{cmemodel}}
\end{figure}

The direct comparison between EIT waves and white-light CMEs revealed that EIT
wave fronts are cospatial with the CME frontal loop \citep{chen09a, dai10}. Such
a result confirmed the theoretical prediction of \citet{chenfa05}, i.e., EIT
waves are the EUV counterparts of the CME frontal loops, whereas the EUV 
extending dimmings are the EUV counterparts of the CME cavity. The cospatiality
implies that the formation mechanism of EIT waves can be directly applied to the
CME frontal loops. Therefore, \citet{chen09a} extended their field-line
stretching model for EIT waves to explain the formation mechanism of the CME
frontal loop. As illustrated by Fig. \ref{cmemodel}, as the core structure,
e.g., a magnetic flux rope, erupts, the resulting perturbation propagates
outward in every direction, with a probability of forming a piston-driven shock
as indicated by the pink lines. However, different from a pressure pulse,
the erupting flux rope keeps pushing the overlying magnetic field
lines to expand, so that the field lines are stretched outward one
by one. For each field line, the stretching starts from the top, e.g.,
point A for the first magnetic line, and then is transferred down to the
leg (point D) with the Alfv\'en speed, by which the first field line is
stretched entirely. The deformation at point A is also transferred
upward to point B of the second magnetic field line with the fast-mode
wave speed. Such a deformation would also be transferred down to its leg
(point E) with the local Alfv\'en speed, by which the entire second
magnetic field line is stretched up. The stretching of the magnetic
field lines compresses the coronal plasma on the outer side of the field
line, producing density enhancements. All the newly formed density
enhancements at a given time form a pattern ({\it green}), which is
observed as the CME frontal loop.

According to this model, the horizontal velocity of the CME footpoints is
$\sim1/3$ of the local fast-mode wave speed ($v_f$), and the radial velocity
of the CME leading loop, i.e., the generally called CME velocity, is equal to
the local fast-mode wave speed, which is several times faster than the plasma
bulk velocity in the CME. Only when the local $v_f$ decreases
below the bulk velocity, the CME becomes a real mass motion, which may happen
at several solar radii. Besides, as noted by \citet{chen11}, this model might
be applied to most CMEs. However, for some blowout CMEs with a very small
velocity, their motion might be a mass motion from the very beginning.

\section{Prospects}

The controversies on ``EIT waves" result mainly from the low cadence of the
observations in the past decade. With the launch of {\it SDO} mission in 2010,
the high-cadence (12 s) observations are unveiling the secret of ``EIT waves"
gradually \citep{liu10, chenwu11}. At the same time, spectroscopic observations
will be of great help \citep{chen10, harr11}.

\section*{Acknowledgements}

PFC is grateful to the SOC members for the invitation to present the review
paper and to the referee for reading the manuscript. This work was
financially supported by Chinese foundations 2011CB811402 and NSFC (11025314,
10878002, and 10933003).


\label{lastpage}
\end{document}